\documentstyle{mn}

\title{Variability in active galactic nuclei: confrontation of
 models with observations}
\author[M. R. S. Hawkins]
       {M. R. S. Hawkins \\
        University of Edinburgh, Royal Observatory, Blackford Hill,
        Edinburgh EH9 3HJ, Scotland, UK}
\date{Accepted ??.
      Received ??;
      in original form ??}

\pagerange{\pageref{firstpage}--\pageref{lastpage}}
\pubyear{2001}

\begin{document}
\label{lastpage}

\maketitle

\label{firstpage}

\begin{abstract}

The variability of active galactic nuclei (AGN) has long held the
promise of shedding light on their detailed structure, and possibly
other astrophysical phenonema.  Different emission mechanisms lead to
different patterns of variability in flux which are in
principle easily distinguishable.  Recent predictions for the expected
spectrum of variations for various models are now in such a form that
they can be compared with the observed statistical properties of AGN
light curves from large scale monitoring programmes.  In this paper, we
use the results of a long term monitoring programme of a large sample
of quasars and Seyfert galaxies, as well as individual light curves
from the literature, to distinguish between the various model
predictions.  The results favour a model based on accretion disc
instability over the starburst model where the variation comes from
a succession of supernova bursts, but it also appears that much of
the observed variation in quasars is due to gravitational microlensing.

\end{abstract}


\section{Introduction}

Forty years after the discovery of quasars there is still much
uncertainty about their structure.  There is, arguably, broad
agreement about the basic nature and overall arrangement of the
various components of a `unified model' for active galactic nuclei
(AGN) \cite{a93}, but the details
have proved very hard to tie down.  In particular, the nature of the
central engine and the radiative transfer processes are the subject
of much debate.  Part of the problem is the paucity of observations
which can effectively distinguish one model from another.  Two types
of constraint which are of particular interest are the spectral energy
distribution and the observed variations in flux.  Although one
can deduce much about AGN structure from properties such as the
optical/xray flux ratio \cite{g91} or the so-called `big blue bump'
\cite{g94}, such measurements are not sufficient to distinguish
between competing models, let alone refine model parameters.

Variations in flux were detected in quasars shortly after their
discovery, and right from the start have played a pivotal role
in constraining quasar morphology.  The early detection of light
fluctuations on a timescale of months provided a fundamental constraint
on the size of the emitting region which has underlain all
subsequent efforts to model the structure of quasars.  In order
to obtain a clearer picture of the emitting regions in AGN
and the associated emission mechanisms, a number of monitoring
programmes have been carried out to measure the spectrum of variations
in different wavebands.

At present there are three basic models for explaining the observed
AGN variability.  The first involves instabilities in the accretion
disc, the central engine powering the energy output \cite{r84}.  The
second postulates that AGN are powered by multiple supernova
explosions or starbursts which result in stochastic variations in
brightness \cite{a94}.  In the third approach, the observed
variations are not intrinsic to the AGN at all, but the result of
gravitational microlensing by small compact bodies or MACHOs along the
line of sight \cite{h93}.  In fact, it seems quite likely that two or
even all three of these processes are present at some level, and so
the task of the observer is to disentangle them all in order to draw
useful astrophysical conclusions.  There is also the possibility that
different mechanisms dominate in different luminosity regimes, and
so in the analysis in this paper we shall divide AGN into two
categories, quasars with $M_{B} < -23$ and Seyfert galaxies with
$M_{B} > -23$.

Although much effort has been put into monitoring quasars
\cite{t94,h94,c96,h96}, both
individually and in samples of various sizes and different selection
criteria, it seems fair to say that little light has so far been
shed on the nature of quasars using these methods.  There appears
to be two main reasons for this.  Firstly, despite the efforts which
have been put into quasar monitoring programmes the data have on the
whole proved inadquate for measuring useful parameters to characterise
the variability.  This is partly because samples of quasars have
been too small, and partly because the run of data has been too short
and inhomogeneous.  As a consequence it has not proved possible to
unambiguously define the fundamental properties of the flux variations
of the quasars.

The second reason that quasar monitoring programmes have not led to
more progress in the understanding of AGN concerns the lack of firm
predictions for variability from the various competing AGN models.
A big step forward in this area has recently been made with the
publication \cite{k98} of a detailed model for AGN variability
from accretion disc instability.  In this paper the authors make
detailed statistical predictions for the spectrum of fluctuations
for their model and also for the starburst model.  These predictions
are presented in a form that enables meaningful comparison with
observations of AGN variability.

It is the purpose of this paper to use the best available observational
data of the variations of AGN to distinguish between the various
models of variability.  To this end we
describe a large scale monitoring programme of a sample containing
some 600 quasars with regularly sampled light curves covering 24
years.  We also use extensive monitoring data for Seyfert galaxies
taken from the recent literature.  Predictions of the models are
published in the form of structure functions and we analyse the
observational data in the same way to provide quantitative comparisons.

\section{The structure function}

In order to extract information from observations of AGN variability
it is necessary to find ways of quantitively characterising the
nature of the variations.  There is a vast literature on time series
analysis, much of it concerned with extracting information from
incomplete or inhomogeneous datasets.  There are several functions
which have been used for analysing AGN variations, including the
structure function, auto-correlation function and Fourier power
spectrum.  Although all of these functions contain very similar
information, in practice they each have advantages in particular
situations.  For example, although for an infinite and complete run
of data the auto-correlation function and power spectrum are
essentially equivalent to each other, for finite runs of data there
are significant differences.  For long runs of evenly spaced data
the power spectrum is to be preferred as it is on the whole easier to
interpret and understand the errors.  For short or inhomogeneous
datasets the auto-correlation function provides a more stable
measurement, but as the individual points are not independent of each
other there can be difficulties with interpretation.  The structure
function is very similar to the auto-correlation function and has
been widely used in the analysis of quasar light curves
\cite{t94,h94,c96,a97a} and microlensing statistics {\cite{w01}.  The
function of choice will depend upon a number of factors, not least of
which is the form in which model predictions have been made.

The present paper has been largely prompted by the publication
\cite{k98} of quantitative predictions for the statistics of AGN
variability, and these are presented in the form of structure
functions.  Accordingly, we shall proceed with the analysis of the
observations in the same way.  The structure function $S$ may be
defined by

\[
S(\tau) =\sqrt{\frac{1}{N(\tau)}\sum_{i<j}[m(t_{j})-m(t_{i})]^{2}}
\]

\noindent
where $m(t_{i})$ is the magnitude measure at epoch $t_{i}$, and the
sum runs over the $N(\tau)$ epochs for which $t_{j}-t_{i} = \tau$.
The interpretation of the structure function in the sense of
identifying specific characteristics of the variation is not usually
feasible.  However, particular models of variability can be shown to
produce structure functions with measurable parametric forms.  Although
in some cases efforts have been made to predict the shape of the
structure function \cite{c97}, it is perhaps more useful to generate
structure functions from simulated data \cite{k98}, and this is the
approach underlying the present paper. In addition to the standard
structure function defined above, we shall for the purpose of
measuring asymmetries also make use of two modified structure
functions $S_{+}$ and $S_{-}$.  These are defined as for $S$ except
that for $S_{+}$ the integration only includes pairs of magnitudes
for which the flux becomes brighter, and for $S_{-}$ for which it
becomes fainter.

\section{Models of AGN variability}

\begin{figure*}
\begin{picture} (560,140) (0,0)
\includegraphics{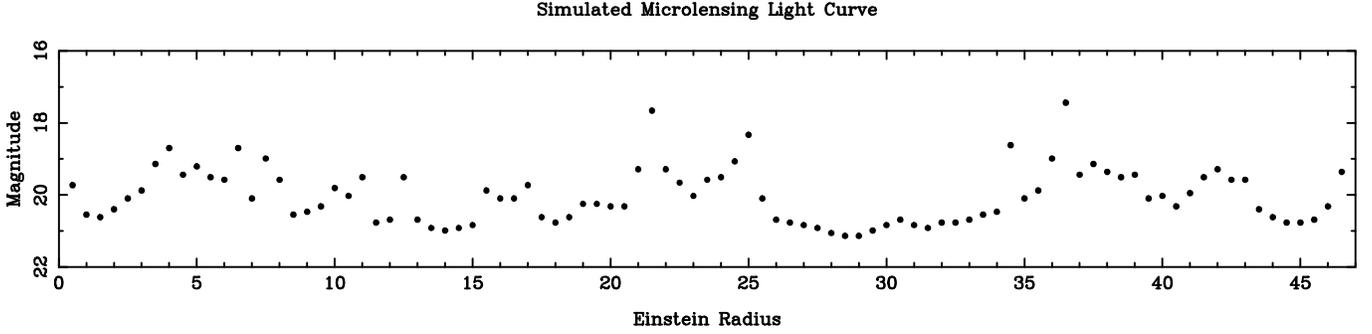}
\end{picture}
\caption{Simulated microlensing light curves from Fig. 2(a) of Lewis
 et al. (1993).
 \label{fig:fig1}}
\end{figure*}

\begin{figure*}
\begin{picture} (560,280) (0,0)
\includegraphics{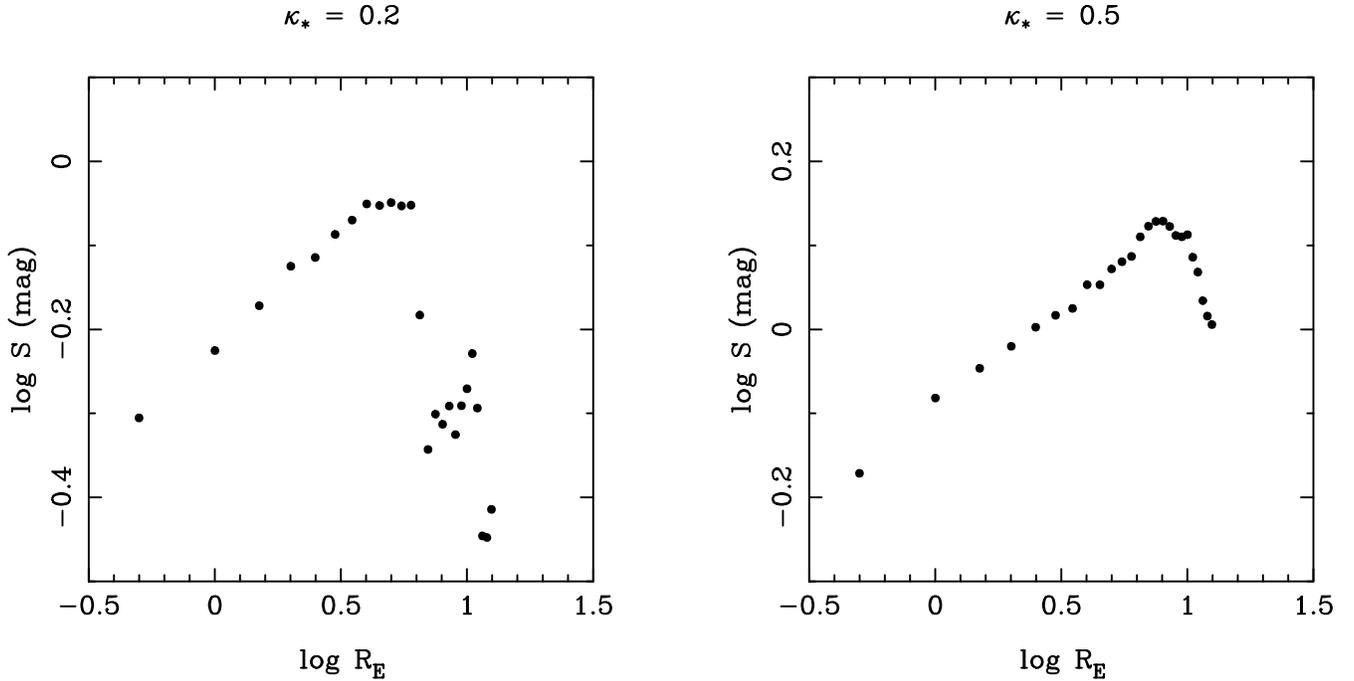}
\end{picture}
\caption{Structure functions for simulated microlensing light curves
 for model $b$ of Schneider \& Weiss (1987) (left hand panel), and
 from Fig. 2(b) of Lewis et al. (1993) (right hand panel).
 \label{fig:fig2}}
\end{figure*}

The mechanism behind AGN variability has been the subject of much
debate since their first discovery.  At present there are three broad
approaches to explaining the observed variations.  The most favoured
model has invoked instabilities in the accretion disc surrounding the
central black hole, but a contending idea is that the variations are
caused by some intermittent sequence of discrete events such as
supernova bursts.  A third possibility is that we are seeing the
effects of gravitational microlensing by a population of small
compact bodies along the line of sight.  Each of these models has
good arguments in its favour, but also fails to explain some aspects
of AGN variability, and it seems likely that all three processes
contribute at some level to the observed light curves with perhaps
one mechanism dominating in any particular regime.  Until recently,
there has been little attempt in the literature to make model
predictions which can easily be tested against observations, but that
is now changing.  In the remainder of this section we review the
current position for the three basic models mentioned above.

\subsection{Disk instability model}

The description of the accretion disc model as proposed by Rees
(1984) contains much discussion about timescales of variability,
but makes no predictions about the spectrum of variations.  Since
then there has been much work on variability in accretion discs
\cite{w92}, but the relation between the various types of instability
and changes in emitted flux have proved hard to tie down and turn
into specific testable predictions for an observed spectrum of
variations.  In fact to the extent that predictions are possible,
they would appear to be at variance with much of the observed
variability.

A major step forward has been made in a recent paper by Kawaguchi
et al. (1998) who develop the cellular-automaton model for disc
instability of Mineshige et al. (1994) to produce quantitative
testable predictions for the shape of the structure function for the
resulting photometric variations of the disc.
The basic idea of the cellular-automaton model is that as matter
flows through an accretion disc it causes instabilities which produce
a spectrum of avalanches of different sizes, which in turn manifest
themselves as variations in emitted flux.  The timescale of the light
variation corresponds to the accretion timescale $\tau_{acc}$, which
is defined \cite{k98} as

\[
\tau_{acc} = 160\left(\frac{r}{10^{2}r_{g}}\right)^{3/2}
 \left(\frac{M}{10^{9}M_{\odot}}\right) \rm{days}
\]

\noindent
where $r_{g}$ is the Schwarzschild radius.  For plausible values of
the black hole mass $M$ this implies a timescale of days or less for
emission from the innermost stable orbit.  However, observations of
quasar light curves imply timescales of at least several years
\cite{h94,h96,c96} which for the cellular-automaton model would mean
emission from a characteristic radial distance of $\sim 1000 r_{g}$.
In the notation of the paper \cite{k98} this would imply
$r_{in}/r_{out} \approx 1000$.  The authors use Monte Carlo
simulations of the model for several parameter sets to produce
simulated light curves from which they can evaluate structure
functions.  The structure functions have the form of a power law
which flattens at the timescale $\tau_{acc}$.  It may be seen from
their Table 2 that the logarithmic slope of their structure function
is not sensitive to the ratio $r_{in}/r_{out}$ or to their other
input parameter $m^{\prime}/m$, the ratio of diffusion mass to inflow
mass, but lies close to $0.44 \pm 0.03$.  This is a stable figure
which can thus be compared with observations.

\subsection{Starburst model}

\begin{figure}
\begin{picture} (280,280) (0,0)
\includegraphics{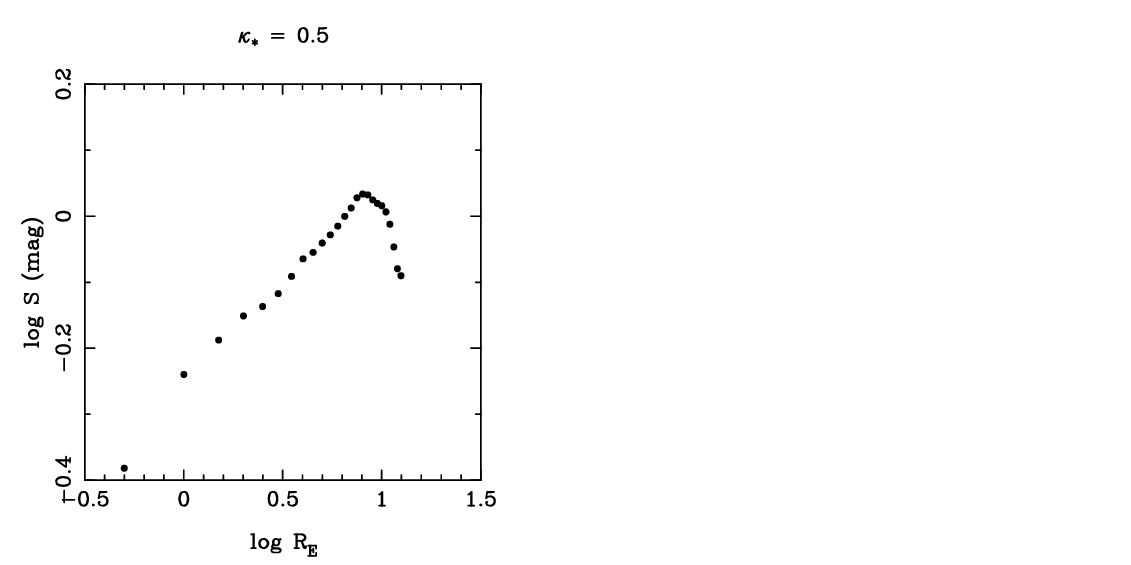}
\end{picture}
\caption{Structure function for the simulated microlensing light
curve from Fig. 2(b) of Lewis et al. (1993).
 \label{fig:fig3}}
\end{figure}

The idea that AGN variability might be caused by a series of
discrete outbursts such as supernova explosions superposed as a
Poisson process, has a long history.  It was however recognised
early on that there was a fundamental problem with the idea which
had come to be known as the `Christmas tree' model \cite{p83}.
The model predicts a relation between the luminosity and amplitude
of quasars which is not observed, and so was ruled out early on as
a serious contender for explaining their structure and variability.
Despite this the model has some attractive features, especially for
explaining low luminosity AGN such as Seyfert galaxies where the
nuclear luminosity could plausibly be explained by a few supernovae
per year, which could at the same time account for the observed
amplitude of variability.  For quasars of even moderate luminosity the
required supernova rate of around one per day is such that the
amplitude of variation would be far too small to be consistent with
observations. Notwithstanding this, Terlevich et al. (1992) made a good
case that the spectroscopic properties of AGN could be explained as
originating from supernovae with their `starburst' model.  The case
that the variability of quasars can be explained \cite{a94} is not
convincing, but it cannot be dismissed as a possible dominating
feature of Seyfert galaxies.

One positive feature of the starburst model is the feasibility of
predicting the spectrum of variations which will be observed.
Kawaguchi et al. (1998), in addition to the accretion disc model,
also produced model light curves and structure functions for starbusts.
They used the formalism of Aretxaga \& Terlevich (1994) to construct
a model and perform Monte Carlo simulations.  Their Fig. 3 shows
model light curves for a quasar of moderate luminosity which
well illustrates the fundamental problem with the starburst model
as applied to quasars.  The amplitude of variation is about 0.2
magnitudes which is far less than would be expected for such an
object \cite{h00}.

The structure functions of the model starburst light curves have a
power law form with a flattening at around 100 days, depending on
model parameters.  The power law section of the structure function
has a logarithmic slope of about $0.83 \pm 0.08$, much larger than
for the accretion disk, and a potential discriminant between the two
models.

\subsection{Microlensing model}

\begin{figure*}
\begin{picture} (560,595) (0,200)
\includegraphics{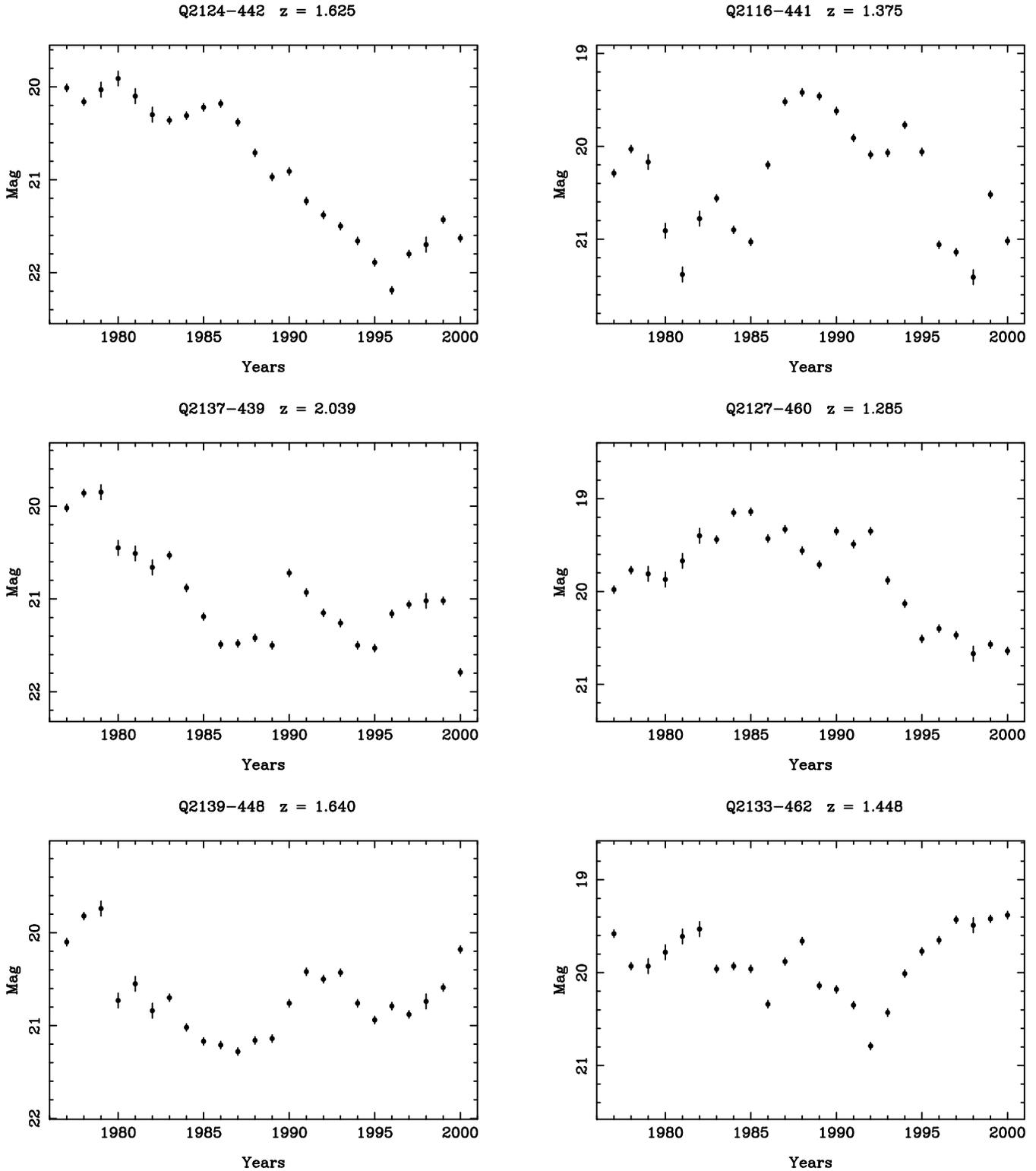}
\end{picture}
\caption{Light curves from the survey of Hawkins (1996), illustrating
 typical characteristics for quasars.
 \label{fig:fig4}}
\end{figure*}

The third broad approach to explaining observed AGN variability
invokes gravitational microlensing.  Unlike the two models already
described, no claim can be made that microlensing explains all
variations seen in AGN.  The hypothesis is that
for quasars the observed long term variations (on a timescale of
several years) are dominated by the effects of microlensing.  Here,
the main interest centres on the microlensing bodies, although there
is the potential to learn about the strucure of the central region
of quasars from the microlensing process.

The idea \cite{h93,h96} is that the dark matter is made up of a
large population of planetary mass compact bodies or MACHOs which
behave as cold dark matter and randomly cross the line of sight to
any particular quasar.  The optical depth to microlensing is such that
on average a quasar's light is lensed by several bodies at any one
time, resulting in a distribution of caustics which causes a
complicated amplification pattern on a timescale of a few years.
Low redshift Seyfert galaxies are too nearby for any significant
probability of microlensing, and so any variation seen in them
must be intrinsic.  Microlensing is well-known to occur in multiply
lensed quasar systems where some variations are only seen in one
image, and on long timescales appear to dominate over the fluctuations
seen in all images which must be intrinsic to the quasar \cite{p98}.

Over the last 20 years or so a number of groups have published
computer simulations of the light curves produced by microlensing
\cite{k86,s87,l93}.  Fig.~\ref{fig:fig1} shows data from Fig. 2(a) of
Lewis et al. (1993) of magnitude versus Einstein radius $R_{E}$,
sampled to enable comparison with observed
light curves.  Structure functions were calculated for these data,
and also for the light curves for model $b$ from Schneider \& Weiss
(1987).  The two simulations are for point sources, but assume
different surface mass density $\kappa_{*}$ which is essentially
equivalent to the optical depth to microlensing \cite{s87}.  The two
structure functions are plotted in the left \cite{s87} and right
\cite{l93} panels of Fig.~\ref{fig:fig2}.  The structure functions
have logarithmic slopes of 0.28 and 0.23 respectively over the linear
part, and there appears to be more power for greater optical depth.

In microlensing simulations, the source size has a very noticeable
effect on the appearance of the light curves, which have a smoother
more rounded structure as it becomes larger.  Fig.~\ref{fig:fig3}
shows the structure function for data from Fig. 2(b) of
Lewis et al. (1993).  These data come from the same distribution
of lenses as for their Fig.2(b), but with a
source size of 0.2 $R_{E}$.  The logarithmic slope of the linear
part of the structure function in Fig.~\ref{fig:fig3} is 0.31.
This is slightly more than for a point source, as is to be expected
from the loss of high frequency components attributable to a resolved
source, although there is only a small difference in amplitude at
long timescales.

\subsection{Comparison of model predictions}

\begin{figure}
\begin{picture} (280,280) (0,0)
\includegraphics{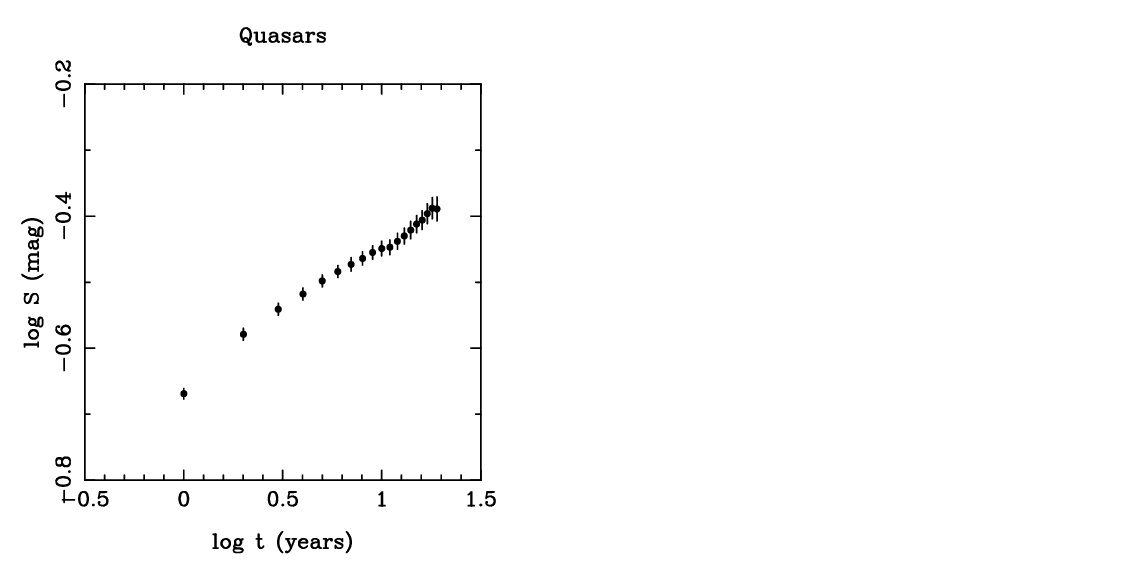}
\end{picture}
\caption{Structure function for the light curves of a sample of
 401 quasars from the survey of Hawkins (1996).
 \label{fig:fig5}}
\end{figure}

It is interesting that the structure functions for all three models
discussed above have a similar morphology.  There is a power law
or logarithmically linear rise from the shortest timescales to
an eventual long timescale break or turnover.  In each model the
power or amplitude of the structure function is somewhat dependent
on the choice of model parameters, but broadly speaking appears
to be less for disc instability than for microlensing, and very much
less for starburst.  The timescales are difficult to compare, as
it is only the starburst model which makes predictions for a
specific timescale, linked to the supernova cooling timescale which
is well known, a characteristic value being 280 days \cite{k98}.  The
accretion disc model is defined in terms of a time step which is left
as a free parameter.  Its value is related to the accretion timescale,
which in turn depends on the size of the emitting region of the
accretion disc.  The timescale of microlensing models depends on
the Einstein radius of the lenses and their mean transverse velocity.
Although the second parameter can be estimated with reasonable
confidence, $R_{E}$ is directly related to the mass of the lenses
which in general is completely unknown.

It is remarkable that the logarithmic slopes of the linear part of
the structure functions show only small dispersions for different
choices of parameter within each of
the three models described above, but that the means are well
separated.  The slopes are $0.83 \pm 0.08$, $0.44 \pm 0.03$ and
$0.25 \pm 0.03$ for the starburst, accretion disc and microlensing
models respectively, which makes for a good opportunity for
distingishing between the models by comparison with observations. 

\section{Observed structure functions}

\begin{figure*}
\begin{picture} (560,280) (0,0)
\includegraphics{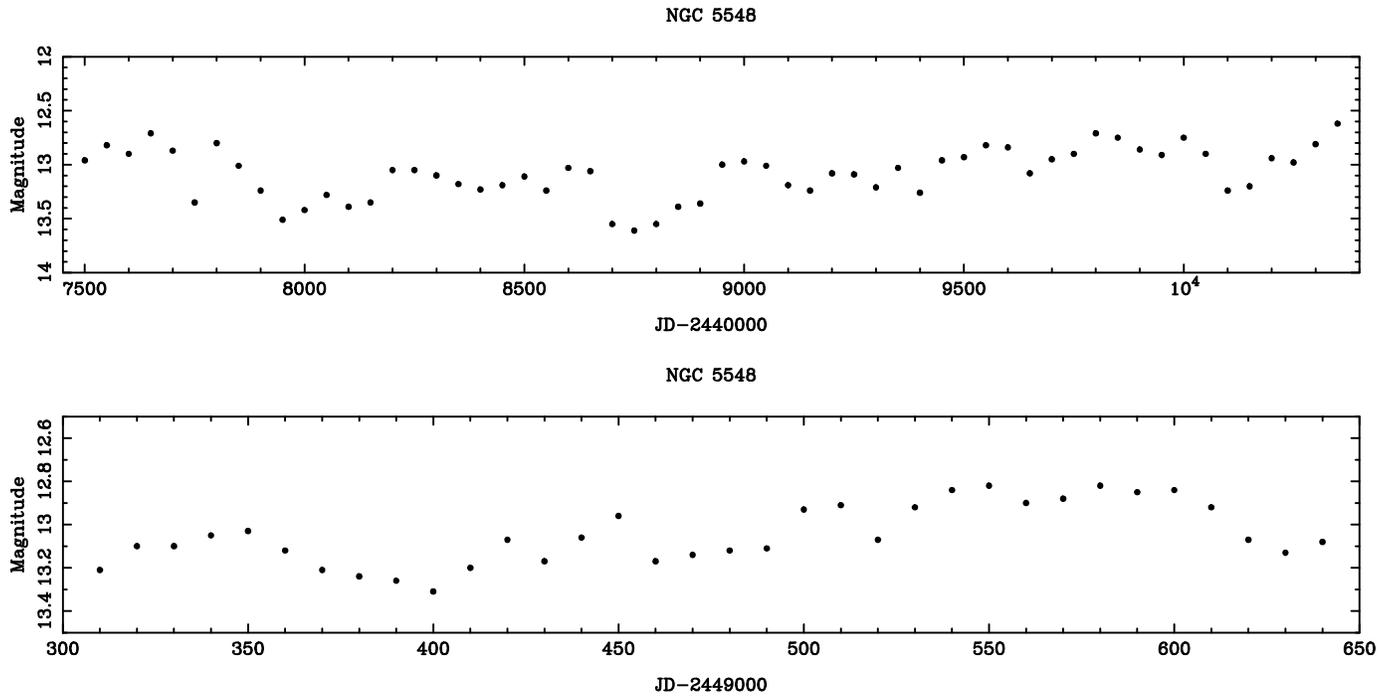}
\end{picture}
\caption{The light curve for NGC 5548 from Peterson et al. (1999)
 from December 1989 till November 1996 (top panel) and from November
 1993 till October 1994 (bottom panel).
 \label{fig:fig6}}
\end{figure*}

\begin{figure*}
\begin{picture} (560,280) (0,0)
\includegraphics{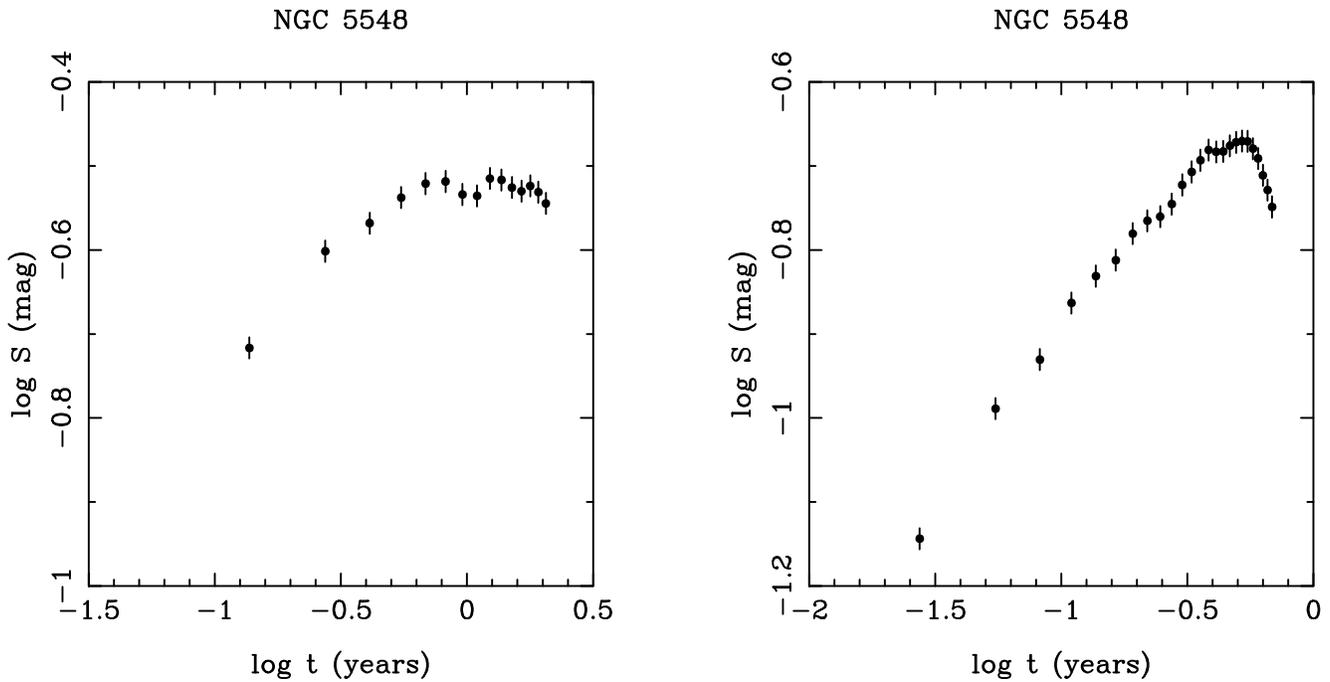}
\end{picture}
\caption{Structure functions for the light curve of NGC 5548 from
 Peterson et al. (1999) and references therein for 50 day intervals
 (left panel) and 10 day intervals (right panel).
 \label{fig:fig7}}
\end{figure*}

\subsection{Quasars}

\begin{figure*}
\begin{picture} (560,395) (0,395)
\includegraphics{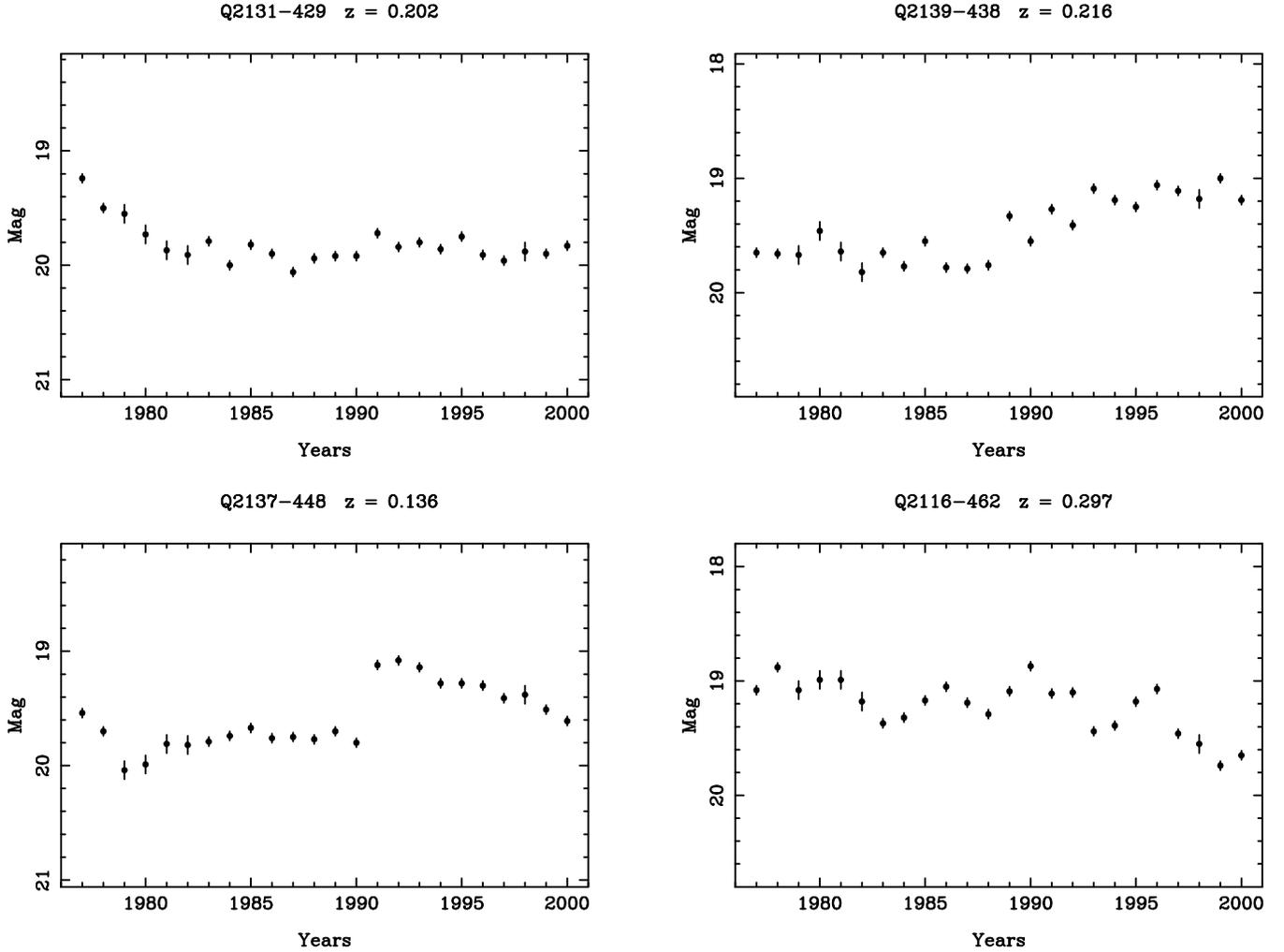}
\end{picture}
\caption{Light curves from the survey of Hawkins (1996), illustrating
 typical characteristics for low redshift Seyfert galaxies.
 \label{fig:fig8}}
\end{figure*}

Over the past ten years or so there have been several major projects
to monitor the light variations of quasars on a timescale of ten years
or more \cite{t94,h94,h96,c96}.  The data for these monitoring
programmes have been presented in the form of structure functions or
auto-correlation functions, and used to look for correlations of
redshift and luminosity with various parameters related to amplitude
and timescale.  It seems fair to say that the search for a correlation
with the amplitude or power of the sructure function has been a lot
more successful than questions involving timescale of variation.

The problem with the existing data is that the structure functions are
not sufficiently well defined to compare with or distinguish between
model predictions.  There are a number of reasons for this.  Perhaps
the most serious difficulty is the smallness of the timespan over
which the quasars have been monitored, which means that features
in the structure functions such as breaks or slopes are not defined
with sufficient accuracy.  A related problem concerns lack of
regularity in the observations.  Unevenly sampled light curves are
notoriously difficult to analyse, resulting in various types of
aliassing and artefacts which can be very misleading in their
interpretation.  A different problem concerns sample sizes.  The
difficulty of monitoring a large number of quasars individually on
a regular basis is such that most monitoring projects have been
based on wide field surveys, especially Schmidt telescopes.  Even
so most samples of quasars have been limited by available redshifts
to around 300 members or less.  A final limitation has concerned the
analysis procedures used, which have been tailored to the unevenly
spaced data that is hitherto all that has been available.

In this paper we present data which go a long way to rectifying
the problems mentioned above.  The light curves are part of a
monitoring programme which has been in progress since 1975,
based on repeated measures of a single UK 1.2m Schmidt telescope
field (ESO/SERC Field 287) centred on 21h 28m -45$^{\circ}$.
The field has been observed several times every year in the $B_{J}$
passband (Kodak IIIa-J emulsion with a Schott GG395 filter) since
1977, as well as on shorter timescales of months, weeks and days,
as well as in other passbands, notably $R$ and $U$.  In this paper
we concentrate on the measures in $B_{J}$ which form an unbroken
sequence of 24 yearly measures from 1977 to 2000.  At least one deep
exposure was obtained every year during this period, and in most years
the measures are based on four or more exposures.  The plates were
measured on the COSMOS or SuperCOSMOS automated measuring machines
at the University of Edinburgh to produce yearly magnitudes for the
$\sim 200,000$ objects in the central 19 square degrees of Field 287.
The photometric error on a measurement from a single plate was
approximately 0.08 mag, and for most years when four plates were
available this reduced to 0.04 mag.  More basic details of the survey
including a discusion of measurement errors have already been published
\cite{h96,h00}, and here we concentrate on updating aspects of the
survey relevant to the present paper.

The quasars in the field have been detected using a number of
techniques including ultra-violet excess, red drop-out, variability,
objective prism searches and radio surveys.  There are estimated to
about 1500 quasars in the field to a magnitude limit of $B_{J} = 22$
of which 610 have been confirmed with redshifts.  Of these, several
complete samples have been compiled according to various well-defined
criteria \cite{h00}.

The light curves show a variety of features, especially on a timescale
of several years or more.  Fig.~\ref{fig:fig4} illustrates some
examples.  For most years the plotted magnitudes are the mean of four
measures, and the error bars are based on the average photometric
errors and the number of plates available in any particular year,
which in most cases was four.  The light curves show no obvious
assymetries in time, or easily definable morphological characteristics.
Perhaps the one thing that can be said is that there appears to be
more power on longer timescales, and that more time is needed to be
sure that a characteristic timescale has been found.

Fig.~\ref{fig:fig5} shows the structure function for the quasar
sample.  In order to make a clear distinction between Seyfert
galaxies which are discussed below, a limit of $z > 0.5$ and
$M_{B} < -23$ was set for the quasars, yielding a sample of 401
objects.  The errors were derived by splitting the sample into 
sub-units and measuring the dispersion in the associated structure
functions.  The structure function in Fig.~\ref{fig:fig5} has a near
power law form, with a logarithmic slope of $0.20 \pm 0.01$.  No
allowance has been made for any time dilation effect since this will
not affect the power law index \cite{k98}, and it is easy to verify
that in a narrow redshift range the index remains the same, albeit
with a somewhat larger uncertainty due to the smaller sample size.

There is no indication of a turn-over in the quasar structure
function at long timescales, which confirms the impression from the
light curves in Fig.~\ref{fig:fig4} that the quasars have not
been monitored for long enough to reach a characteristic timescale.
Equally there is no sign of the observations becoming dominated by
noise at short timescales.  This again confirms an impression from
Fig.~\ref{fig:fig4} that the coherent variations are much larger than
the errors on the observations.

\subsection{Seyfert galaxies}

\begin{figure}
\begin{picture} (280,280) (0,0)
\includegraphics{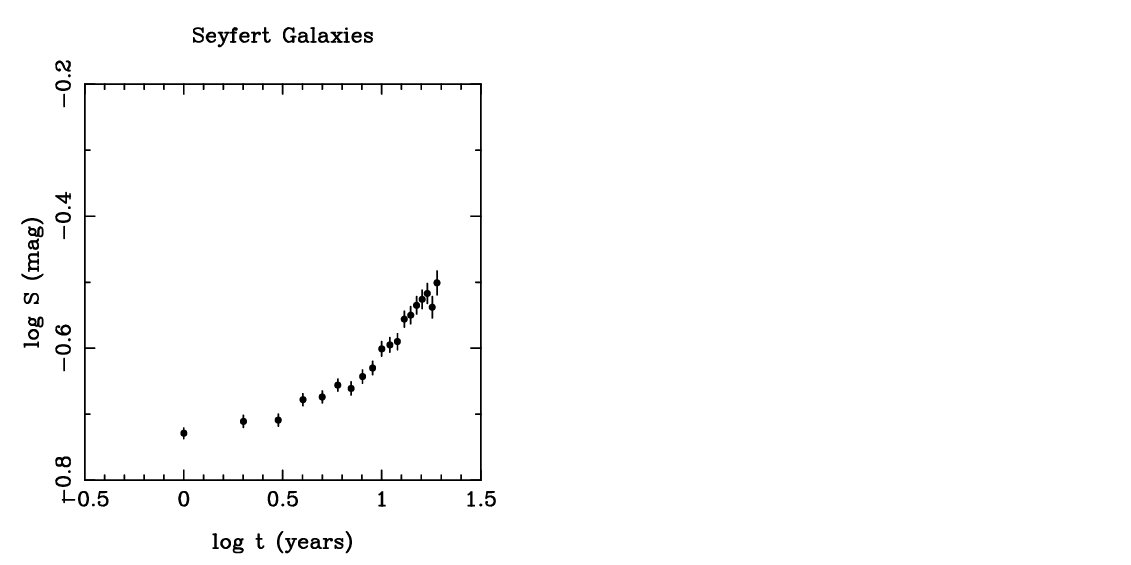}
\end{picture}
\caption{Structure function for the light curves of a sample of
 45 Seyfert galaxies from the survey of Hawkins (1996).
 \label{fig:fig9}}
\end{figure}

Seyfert galaxies are well known to vary in brightness on a timescale
of months or less which makes them relatively easy to monitor within
the constraints of modern policies for the allocation of telescope
time.  However, the project to monitor NGC 5548 \cite{p99} has
surpassed all other efforts.  Fig.~\ref{fig:fig6} (top panel) shows
the whole light curve sampled every 50 days for the purpose of
measuring the structure function.  The bottom panel shows the light
curve from a particularly well observed period in 1993/4, sampled
every 10 days.  Error bars are not plotted as the errors are
approximately the same size as the points of the plots.

The length and frequency of observation of this light curve make it
ideal for evaluating the structure function.  The observations were
made with spectrographs on a number of telescopes as spectra were
essential to the investigators' main programme.  The photometric
observations were obtained by defining a window in the spectrum
close to the $V$ band and integrating within it.  For most of the
observations the errors are about 2\% (0.02 magnitudes), but sometimes
rise to twice this, about as large as the points in
Fig.~\ref{fig:fig6}.

The structure functions for the two light curves in Fig.~\ref{fig:fig6}
are shown in Fig.~\ref{fig:fig7}.  Errors were estimated by re-sampling
the data in various ways and measuring the resulting dispersion
in the structure functions.  The left hand panel shows a rise
to longer timescales with a gradually decreasing slope, becoming
flat at a timescale of about a year.  The logarithmic slope does not
become truly linear, and so to investigate the behaviour on short
timescales we refer to the right hand panel for data sampled every
10 days.  Here there is a well defined power law relation with
logarithmic slope $0.38 \pm 0.01$ with an eventual flattening at
longer timescales.

The survey \cite{h96} described in the previous section contains,
in addition to the quasars, a number of Seyfert galaxies.  Here again,
in order to make a clear distinction with the quasar sample we
define our Seyfert galaxy sample with $M_{B} > -23$ and $z < 0.3$
The light curves for this sample of 45 members have a very different
character to those of the quasars.  Fig.~\ref{fig:fig8} shows some
typical examples, with errors and other details as for
Fig.~\ref{fig:fig4}.  On the whole the light curves have small
amplitudes with variations characterised by small fluctuations on
short timescales, and smooth long term gradients of apparently modest
amplitude.

The structure function for this sample is shown in Fig.~\ref{fig:fig9}
with errors calculated as for Fig.~\ref{fig:fig5}.  Again the data
for Seyfert galaxies show a very different character from that for
the quasars.  The amplitude is much smaller as would be expected
from the appearance of the light curves.  There is a linear section
of the plot which has a slope of $0.36 \pm 0.02$, close to that of
NGC 5548, and much larger than the figure of $0.20 \pm 0.01$ for
the quasars.  The structure function flattens at short timescales.
The reason for this is not immediately apparent, but may be related
to measurement errors.

\section{Time asymmetries in structure functions}

\begin{figure*}
\begin{picture} (560,280) (0,0)
\includegraphics{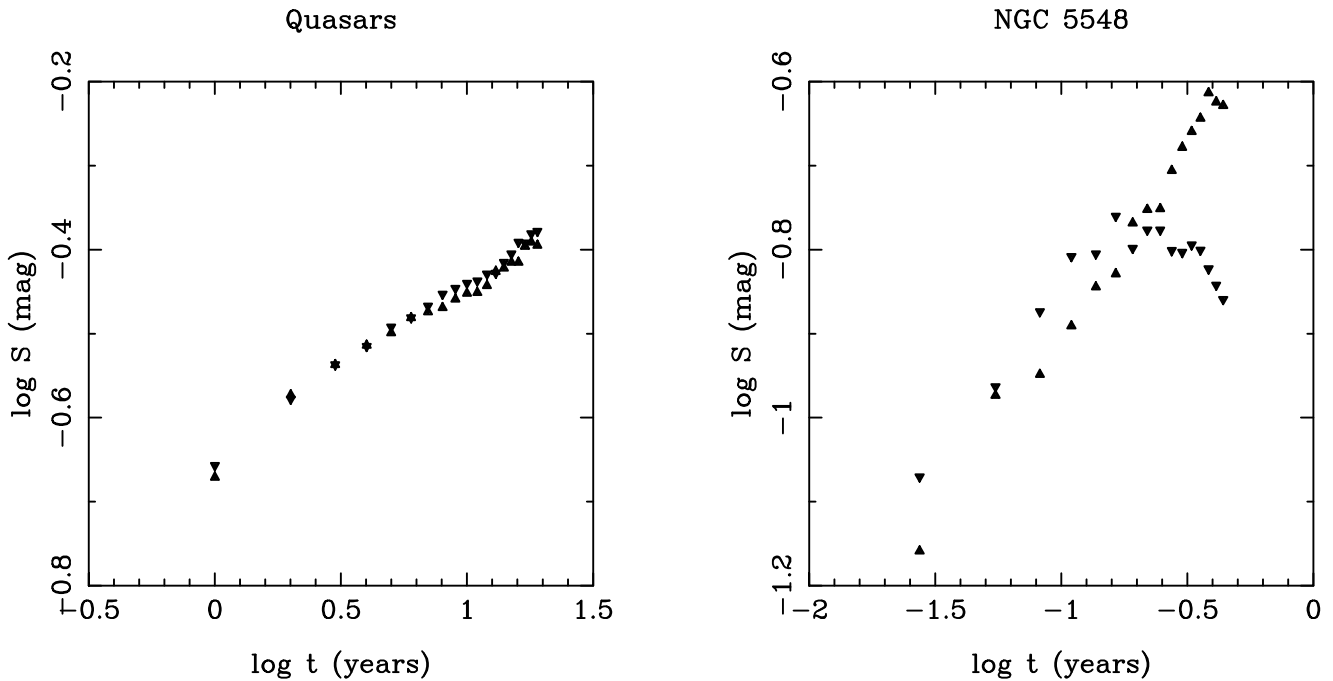}
\end{picture}
\caption{Time asymmetrical structure functions for the data in
 Fig.~\ref{fig:fig5} (left panel) Fig.~\ref{fig:fig7} (right panel).
 Functions for increasing ($S_{+}$) and decreasing ($S_{-}$) brightness
 are shown by upward and downward pointing triangles respectively.
 \label{fig:fig10}}
\end{figure*}

\subsection{Model predictions}

In addition to the slope of the structure function, Kawaguchi et al.
(1998) discuss a further discriminant between models of AGN
variabiliy using the functions $S_{-}$ and $S_{+}$ defined in section
2 above.  These functions provide a measure of the underlying asymmetry
of the emission process as manifested in the light curves.  It is a
fortuitous circumstance that the three models for variability
discussed above make qualitatively different predictions for the
relationship between $S_{-}$ and $S_{+}$.

The disc instability model as put forward by Kawaguchi et al. (1998)
predicts a spectrum of `avalanches' in the accretion disc which
involve a gradual brightening of the disc followed by a sudden drop
in flux.  This pattern will manifest itself as a statistical asymmetry
in the light curves which can be detected by comparing $S_{-}$
and $S_{+}$.  In this case $S_{-}$ should be larger than $S_{+}$
towards shorter timescales.

By contrast, the starburst model is dominated by the light of
supernovae which are well known to rise rapidly in brightness at
the onset of the explosion after which they gradually fade.  Again,
this asymmetry will be apparent in the AGN light curve as a statistical
asymmetry, but this time such that $S_{+}$ is larger than $S_{-}$
towards shorter timescales \cite{k98}.  In both cases the difference
between $S_{-}$ and $S_{+}$ depends on the model parameters, but
should be measurable with sufficiently good data.

In the case of microlensing, which is an essentially symmetrical
process in the context of quasar variability, it should not be
possible to determine which way time is running by any statistical
tests on the light curves.  $S_{-}$ and $S_{+}$ should be
indistinguishable and lie on top of each other.

\subsection{Observations}

The idea of using time asymmetry in quasar light curves as a
diagnostic for quasar variability has already received some
attention in the literature \cite{h96,a97b}.  The conclusion of
both these studies is that statistically the light curves show no
detectable departure from symmetry.  The data used to produce the
structure functions in Figs.~\ref{fig:fig5} and \ref{fig:fig7} can
also be used to calculate the functions $S_{-}$ and $S_{+}$.
Fig.~\ref{fig:fig10} shows these two functions plotted together
for comparison.  Error bars have been omitted in the interests of
clarity, but are as for the data in Figs.~\ref{fig:fig5} and
\ref{fig:fig7}.

The $S_{-}$ and $S_{+}$ functions for quasars in the left hand panel
lie very close to each other, and are indistinguishable within the
errors.  This implies that the observed variation is symmetrical or
time reversible.  The data for the Seyfert galaxy NGC 5548 on the
other hand show strong and significant differences between the two
functions.  On short timescales $S_{-}$ and $S_{+}$ both have a
logarithmic slope of about 0.4 but $S_{-}$ is systematically larger
than $S_{+}$.  This implies that there is an asymmetry in the light
curve in the sense that the flux increases more slowly than it falls
\cite{k98}.  On a time scale longer than three months the two curves
separate, which is probably due to the limited length of the data run
with 10 day sampling.

\section{Confrontation of models with observations}

We are now ready to test the three models described in section 3,
which we shall do separately for quasars and Seyfert galaxies.
The model predictions for the slope of the structure function are
$0.83 \pm 0.08$, $0.44 \pm 0.03$ and $0.25 \pm 0.03$ for the
starburst, accretion disc and microlensing models respectively. 
These figures are given in Table 1, along with the observed slopes
for comparison.
The slope measured for the quasar sample was $0.20 \pm 0.01$ which
clearly rules out the starburst model.  It also appears to be a long
way from the accretion disc model but is consistent with the
predictions for microlensing.  It is however worth noting that
for resolved sources the microlensing structure function gets steeper,
presumably as high frequency detail is blurred out, and there will
certainly come a point where it is inconsistent with the data.

\begin{table}
\centering
\caption{Structure Function Slopes}
\begin{tabular}{ll} \hline
&\\
\multicolumn{2}{c}{Model Predictions}\\
\hline
Model&Slope\\ \hline
&\\
Starburst       &$0.83\pm0.08$\\
Disc Instability&$0.44\pm0.03$\\
Microlensing    &$0.25\pm0.03$\\ \hline
&\\
\multicolumn{2}{c}{Observations}\\ \hline
AGN Class&Slope\\ \hline
&\\
Seyfert Galaxies&$0.36\pm0.02$\\
Quasars         &$0.20\pm0.01$\\ \hline
\end{tabular}
\end{table}

We can further compare the models by testing for time asymmetries in
the light curves.  For short timescales the starburst, accretion disc
and microlensing models predict $S_{-} < S_{+}$, $S_{-} > S_{+}$
and $S_{-} = S_{+}$ respectively.  In fact as we have seen, for
quasars $S_{-}$ and $S_{+}$ are effectively coincident, again
favouring the microlensing model.

With Seyfert galaxies we find a very different picture to quasars.
The slope of the structure function for NGC 5548 and for the
Seyfert galaxy sample can be taken as $0.38 \pm 0.01$ which again
appears to rule out the starburst model, although this time it lies
close to the prediction of the accretion disc model.  The microlensing
model also appears to be excluded for any plausible value of the
source size.  Again we can further test the models by looking for
time asymmetries, and we have found that for Seyfert galaxies
$S_{-} > S_{+}$.  This is in agreement with the prediction for the
accretion disc model, and confirms the result for the structure
function slope.

To summarise, for Seyfert galaxies the observations favour the
accretion disc model, and for quasars the microlensing model. 
It appears that the starburst model is ruled out for all AGN.

\section{Discussion}

Although within the terms of reference of this paper the results in
the previous section appear rather conclusive, there are a number
of caveats which must be made clear.  The predictions for all three
models were based on a very limited coverage of the relevant parameter
spaces, and although in each case the dispersion in the structure
function slope did not appear to be large, and very much less than
the differences between the mean predictions of the models, it is
entirely possible that a more thorough exploration of the parameter
spaces could alter the picture significantly.  No doubt in due course
such data will become available, and the results given here can be
tightened accordingly.

In this paper we have concentrated on the slope of the structure
function because for all three variability mechanisms considered it
appears to be a robust quantity with lttle dependence on the choice
of model parameters.  There are other quantities such as the timescale
of variability or the amplitude of the turnover in the structure
function which could in principle also be used.  The problem here
is the strong dependence on model parameters, or sensitivity to
the window function of the data.  We therefore content ourselves for
the moment with the structure function slope.  However, having once
decided which variability mechanism is in operation, the process
can be reversed to estimate the values of model parameters.

The cellular-automaton model for disc instability \cite{m94} has
been used here as a prototype for accretion disc models because
of the testable predictions which are mow available \cite{k98}.
There is of course much controversy over the details of emission
from an AGN accretion disc, and it may well be that other
accretion disc models make very different predictions for the
logarithmic slope of the structure function.  When testable model
data become available it seems possible that the conclusions of this
paper will have to be modified.  It also goes without saying that
the predictions of most models can be stretched to accommodate data
by a careful choice of parameters, and it may be that the conclusions
of this paper turn out to be less conclusive than they appear to be
at present.

Notwithstanding these caveats, the conclusions reached in section 6
appear to be well supported by the data.  Taking the results at face
value, we find that photometric variation in Seyfert galaxies is best
explained by instabilities in an accretion disc, while for quasars
the dominant mechanism for variability is microlensing.  At first
sight this might seem paradoxical, but in fact the two explanations
sit quite well together.  We can account for them by a model in
which the fluctuations in the light from the accretion disc become
smaller with increasing luminosity, but the effects of microlensing
become more pronounced at higher redshift, which for quasars typically
means higher luminosity.

This model can be tested directly in the rather rare situations
where quasar variations can unambiguously be separated into those
which are intrinsic to the quasar and those which are not.  The
best example for this purpose is the gravitational lens system
0957+561 which is a luminous quasar split into two components by
the lensing effect of an intervening galaxy.  The two images have
been monitored intensely over the last 20 years or so, primarily
for the purpose of determining the time delay between the two
components and hence measuring the value of Hubble's constant
\cite{k97}.  The time delay was measured from the displacement of
features which occured in both light curves and are clearly
intrinsic to the quasar.  These features are on the whole of
short duration ($\sim 100$ days) and small amplitude ($\sim 0.1$
magnitudes) as may be seen from Fig. 4 of Kundi\'{c} et al. (1997).

It has been known for a long time that as well as these intrinsic
variations, the two components of 0957+561 also vary independently on
a timescale of several years with an amplitude greater than 0.3
magnitudes \cite{p98}.  This mode of variation is generally accepted
as microlensing by compact bodies along the line of sight to the
quasar images.  The difference between the two modes of variation
is well illustrated in Fig. 1 of Kundi\'{c} et al. (1997) where
the small repeating features contrast with the larger long term
variation due to microlensing.  These observations support the idea
that for quasars intrinsic variations are small (certainly smaller
than for Seyfert galaxies) and that microlensing can easily dominate
the observed variations.  

\section{Conclusions}

In this paper we have tested the predictions of three models for
variability in AGN against observations.  We
have compared the predicted logarithmic slope of the structure
function for starburst, accretion disc instability and microlensing
models with the observed structure function for Seyfert galaxies and
quasars.  We find that for Seyfert galaxies the starburst and
microlensing models are not compatible with the observed data, but
the accretion disc model is in good agreement with it.
For quasars the starburst and accretion disc instability models
are ruled out while the microlensing model agrees well with the
observations.

We conclude that the variations in AGN are best
explained by a model in which at low luminosities the observed 
variations are caused by instabilities in the accretion disc.
With increasing luminosity this type of variation becomes smaller
in amplitude and the observed variation becomes dominated by the
effects of microlensing.  This increase is associated with the
larger optical depth to microlensing for most quasars.

\section*{Acknowledgments}

I wish to thank members of the Anglo Australian Observatory and
the Wide Field Astronomy Unit of the University of Edinburgh
for their support in obtaining and measuring the Schmidt plates
needed for this project.


\begin{thebibliography}{99}

\bibitem[Antonucci 1993]{a93} Antonucci R., 1993, ARA\&A, 31, 473
\bibitem[Aretxaga \& Terlevich 1994]{a94} Aretxaga I., Terlevich R.,
 1994, MNRAS, 269, 462
\bibitem[Aretxaga et al.,\ 1997]{a97a} Aretxaga I., Cid Fernandes R.,
 Terlevich R., 1997, MNRAS, 286, 271
\bibitem[Aretxaga 1997]{a97b} Aretxaga I., 1997, Rev. Mexicana
 Astron. Astrofis., 6, 207
\bibitem[Cid Fernandes et al.,\ 1997]{c97} Cid Fernandes R.,
 Terlevich R., Aretxaga I., 1997, MNRAS, 289, 318
\bibitem[Cristiani et al.,\ 1996]{c96} Cristiani S., Trentini S.,
 La Franca F., Aretxaga I., Andreani P., Vio R., Gemmo, A., 1996,
 A\&A, 306, 395
\bibitem[George \& Fabian 1991]{g91} George I.M., Fabian A.C.,
 1991, MNRAS, 249, 352
\bibitem[Gondhalekar et al.,\ 1994]{g94} Gondhalekar P.M., Kellet B.J.,
 Pounds K.A., Matthews L., Quenby J.J., 1994, MNRAS, 268, 973
\bibitem[Hawkins 2000]{h00} Hawkins M.R.S., 2000, A\&AS, 143, 465
\bibitem[Hawkins 1996]{h96} Hawkins M.R.S., 1996, MNRAS, 278, 787
\bibitem[Hawkins 1993]{h93} Hawkins M.R.S., 1993, Nat, 366, 242
\bibitem[Hook et al.,\ 1994]{h94} Hook I.M., McMahon R.G., Boyle B.J.,
 Irwin M.J., 1994, MNRAS, 268, 305
\bibitem[Kawaguchi et al.,\ 1998]{k98} Kawaguchi T., Mineshige S.,
 Umemura M., Turner E.L., 1998, ApJ, 504, 671
\bibitem[Kayser et al.,\ 1986]{k86} Kayser R., Refsdal S., Stabell R.,
 1986, A\&A, 166, 36
\bibitem[Kundi\'{c} et al.,\ 1997]{k97} Kundic\'{c} T. et al., 1997,
 ApJ, 482, 75
\bibitem[Lewis et al.,\ 1993]{l93} Lewis G.F., Miralda-Escud\'{e} J.,
 Richardson D.C., Wambsganss J., 1993, MNRAS, 281, 647
\bibitem[Mineshige et al.,\ 1994]{m94} Mineshige S., Ouchi B.,
 Nishimori H., 1994, PASJ, 46, 97
\bibitem[Pelt et al.,\ 1998]{p98} Pelt J., Schild R., Refsdal S.,
 Stabell R., 1998, A\&A, 336, 829
\bibitem[Peterson et al.,\ 1999]{p99} Peterson B.M. et al., 1999,
 ApJ, 510, 659
\bibitem[Pica \& Smith 1983]{p83} Pica A.J., Smith A.G., 1983,
 ApJ, 272, 11
\bibitem[Rees 1984]{r84} Rees M.J., 1984, ARA\&A, 22, 471
\bibitem[Schneider \& Weiss 1987]{s87} Schneider P., Weiss A.,
 1987, A\&A, 171, 49
\bibitem[Terlevich et al.,\ 1992]{t92} Terlevich R., Tenorio-Tagle G.,
 Franco J., Melnick J., 1992, MNRAS, 255, 713
\bibitem[Trevese et al.,\ 1994]{t94} Trevese D., Kron, R.G.,
 Majewski S.R., Bershady M.A., 1994, ApJ, 433, 494
\bibitem[Wallinder et al.,\ 1992]{w92} Wallinder F.H., Kato S.,
 Abramowicz M.A., 1992, A\&AR, 4, 79
\bibitem[Wyithe \& Turner 2001]{w01} Wyithe J.S.B., Turner E.L.,
 2001, MNRAS, 320, 21

\end{thebibliography}
\end{document}